\documentclass[manuscript]{aastex}

\newcommand{\OI}{O\,{\sc i}}
\newcommand{\CI}{C\,{\sc i}}
\newcommand{\CII}{C\,{\sc ii}}

\newcommand{\NII}{N\,{\sc ii}}
\newcommand{\NeII}{Ne\,{\sc ii}}
\newcommand{\NIII}{N\,{\sc iii}}
\newcommand{\OIII}{O\,{\sc iii}}
\newcommand{\Sgra}{Sgr\,A$^*$}

\def\,{\thinspace}

\def\Lsun{{\hbox {$L_\odot$}}} 
 
\def\nh2{{\hbox {$n({\rm H_2})$}}}

\slugcomment{To appear in ApJ Letters}

\shorttitle{Shocks versus Radiation near Sgr\,A$^*$}

\shortauthors{Goicoechea et al.}

\begin{document}

\title{\textit{Herschel}\altaffilmark{1} Far-Infrared Spectroscopy of the Galactic Center.\\
Hot Molecular Gas: Shocks versus Radiation near Sgr\,A$^*$}


\author{Javier R. Goicoechea$^2$, M. Etxaluze$^2$, J. Cernicharo$^2$, M. Gerin$^3$, D. A. Neufeld$^4$,\\ A. Contursi$^5$, T. A. Bell$^2$,
M. De Luca$^3$, P. Encrenaz$^3$, N. Indriolo$^4$,\\ D. C. Lis$^{6}$, E. T. Polehampton$^{7,8}$,
P. Sonnentrucker$^{9}$}
\email{jr.goicoechea@cab.inta-csic.es}


\altaffiltext{1}{\textit{Herschel} is an ESA space observatory with science instruments provided by European-led
Principal Investigator consortia and with important participation from NASA.}

\altaffiltext{2}{Departamento  de Astrof\'{\i}sica. Centro de Astrobiolog\'{\i}a. CSIC-INTA.
Carretera de Ajalvir, Km 4. Torrej\'on de Ardoz, 28850, Madrid, Spain.}

\altaffiltext{3}{LERMA, UMR 8112 du CNRS, Observatoire de Paris, \'École Normale Sup\'erieure, France}

\altaffiltext{4}{The Johns Hopkins University, Baltimore, MD 21218, USA}

\altaffiltext{5}{Max-Planck-Institut f\"ur extraterrestrische Physik (MPE), Postfach 1312, D-85741 Garching, Germany}

\altaffiltext{6}{California Institute of Technology, Pasadena, CA 91125, USA}

\altaffiltext{7}{RAL Space, Rutherford Appleton Laboratory, Chilton, Didcot, Oxfordshire, OX11 0QX, UK}

\altaffiltext{8}{Institute for Space Imaging Science, University of Lethbridge, 4401 University Drive, Lethbridge, Alberta T1J 1B1, Canada}

\altaffiltext{9}{Space Telescope Science Institute, Baltimore, MD 21218, USA}


\begin{abstract}

We present a $\sim$52-671\,$\mu$m spectral scan toward \Sgra~taken~with the PACS 
and SPIRE spectrometers onboard \textit{Herschel}. The achieved angular resolution  
allows us to separate, for the first time at far-IR wavelengths,  the emission 
toward the central cavity (gas in the inner central parsec of the galaxy)
from that of the  surrounding circum-nuclear disk.
The spectrum toward \Sgra~is dominated by  strong 
[\OIII],  [\OI],   [\CII], [\NIII],  [\NII],  and [\CI]   fine structure lines  (in decreasing order of luminosity) 
arising in gas irradiated by UV-photons  from the central stellar cluster. 
In addition, rotationally excited lines of $^{12}$CO
(from $J$=4-3 to 24-23), $^{13}$CO, H$_2$O, OH, H$_3$O$^+$, HCO$^+$ and HCN, 
as well as  ground-state absorption lines of OH$^+$, H$_2$O$^+$, H$_3$O$^+$, 
CH$^+$, H$_2$O, OH, HF, CH and NH are detected.
The excitation of the $^{12}$CO ladder is consistent with a \textit{hot} isothermal 
component at $T_{\rm k} \simeq 10^{3.1}$~K and $n$(H$_2$)$\lesssim$10$^4$\,cm$^{-3}$. 
It is also consistent with a distribution of  temperature components 
at higher density with most CO  at $T_{\rm k}\lesssim$300\,K.
The detected molecular features suggest that, at present, neither very enhanced X-ray, nor cosmic-ray fluxes
play a dominant role in the heating of the \textit{hot} molecular gas.
The   \textit{hot} CO~component   (either the bulk of the CO column  or just a small fraction depending on
the above scenario) results from a combination of  UV-~and shock-driven heating.
If irradiated dense clumps/clouds do not exist, 
shocks likely dominate the heating of the $\textit{hot}$ molecular gas.
This is consistent with the  high-velocity gas detected toward  \Sgra.

\end{abstract}


\clearpage

\section{Introduction}

The Galactic center (GC), in particular the interstellar material in the immediate vicinity  ($<5$~pc) of the 
central black hole, 
represents a unique environment for our understanding of galactic nuclei and  galaxy evolution.
 At a distance $d=8.0\pm0.5$~kpc \citep{Rei93}, the nucleus of our galaxy is a few hundred  times 
closer than the nearest active galactic nuclei (AGNs), 
thus allowing high spatial resolution studies. 
The distribution of gas and dust toward the GC \citep[][]{Gen10}
consists of a central cavity of radius $\sim$1.5\,pc containing
 warm dust and gas heated and ionized by the central cluster of massive stars orbiting
close to the black hole (\Sgra~radio source position).
Some of the ionized gas streamers (the ``mini-spiral'') bring material
close to the very center \citep[][see Figure~1]{Yus87}.

Between $\sim$1.5\,pc and $\sim$5\,pc, a disk of denser molecular gas exists
\citep[the circum-nuclear disk or  CND;][]{Gus87}. However, its density
is not well constrained ($\sim 10^4$-$10^8$\,cm$^{-3}$) and it is not yet clear whether all the material
in the  CND is stable against the strong tidal forces 
in the region or has a more transient nature \citep[][]{Bra05,Mon09,Req12}. 
The molecular gas toward the central
cavity has been less studied, first because 
high angular resolution is required to separate the different components and also because column densities are inevitably
 lower and  emission lines are intrinsically weak.
Nevertheless, the detection of CO ro-vibrational lines in absorption \citep{Mon01} and
of broad NH$_3$~(6,6) emission lines 
close to \Sgra~\citep{Her02} suggests that \textit{hot} molecular gas must exist interior to the CND.

Owing to the lower dust extinction at far-IR wavelengths and
because of the strong emission from the interstellar  component related to AGN and
star formation activity, the relevance of far-IR
spectroscopy to characterize   extragalactic nuclei has notably increased  \citep[\textit{e.g.,}][]{Wer10}.
The far-IR spectrum of the Milky Way's nucleus is obviously a key template. 

The angular resolution achieved by the \textit{Herschel Space Observatory} \citep[$\sim10''-40''$;][]{Pil10} allows us to
separate  the emission of the central cavity from the CND.
In this \textit{Letter} we present initial results from a \textit{Herschel} far--IR spectroscopic study
of the GC. We present the complete
 PACS \citep{Pog10} and SPIRE \citep{Gri10} spectra  toward \Sgra~that are part of the 
PRISMAS and SPECHIS Guaranteed-Time Programs. We discuss  the 
 properties and possible origin of the atomic and of the \textit{hot} molecular gas.

\section{Observations and Data Reduction}

PACS spectra between $\sim$52 and $\sim$190\,$\mu$m were obtained during 2011 March and 2012 March.
The PACS spectrometer   provides 25 spectra  
over a 47$''$$\times$47$''$ field-of-view (FoV) resolved in 5$\times$5  ``spaxels'', 
each with a size of $\sim$9.4$''$. 
The resolving power varies between $R$=$\lambda$/$\Delta\lambda$$\simeq$1000 at $\sim$100\,$\mu$m
 and $R$$\simeq$5000 at $\sim$70\,$\mu$m.
The central spaxel was centered on
\Sgra~($\alpha_{2000}$: 17$^h$45$^m$40.04$^s$, $\delta_{2000}$: $-29$$^o$00$'$28.19$''$). 
The observations were carried out in the  ``unchopped'' mode (IDs 1342217802 and 342242442).
Background subtraction was achieved by removing 
the telescope background spectrum measured on a
distant  off­-position. 
The observing time was $\sim$2.4\,hr. 
The measured width of the point spread function 
is relatively constant for
$\lambda$$\lesssim$100\,$\mu$m ($\sim$spaxel size) but increases at longer wavelengths.
In particular only $\simeq$40\%  of a point 
source emission would fall in the central spaxel at $\simeq$190\,$\mu$m.
Therefore, owing to the extended nature of the emission, the flux measured by a single spaxel cannot be used
individually.
Instead, we added the 3$\times$3 central spaxels ($\sim$30$''$$\times$30$''$).

SPIRE-FTS observations between $\sim$194 and $\sim$671\,$\mu$m
were obtained during  2011 February  (ID1342214845). 
The SPIRE-FTS uses two bolometer arrays covering 
the 194-313\,$\mu$m  
and 303-671\,$\mu$m 
bands at 0.04\,cm$^{-1}$ resolution ($R$$\simeq$500-1000). 
The two arrays contain 19  and 37  
detectors separated by 
$\sim$2 beams (51$''$ and 33$''$ respectively). The unvignetted FoV is $\sim$2$'$.
The observing time was 798\,s.
The \textit{Herschel} data were processed with HIPE\,9.2. 
 Table~\ref{tbl-fluxes} summarizes the atomic and
 CO  line intensities obtained in a
$\sim$30$''$$\times$30$''$ aperture.

\section{Results: Spectroscopy}

\begin{figure*}[ht]
\centering %
\includegraphics[width=1\textwidth,angle=-0]{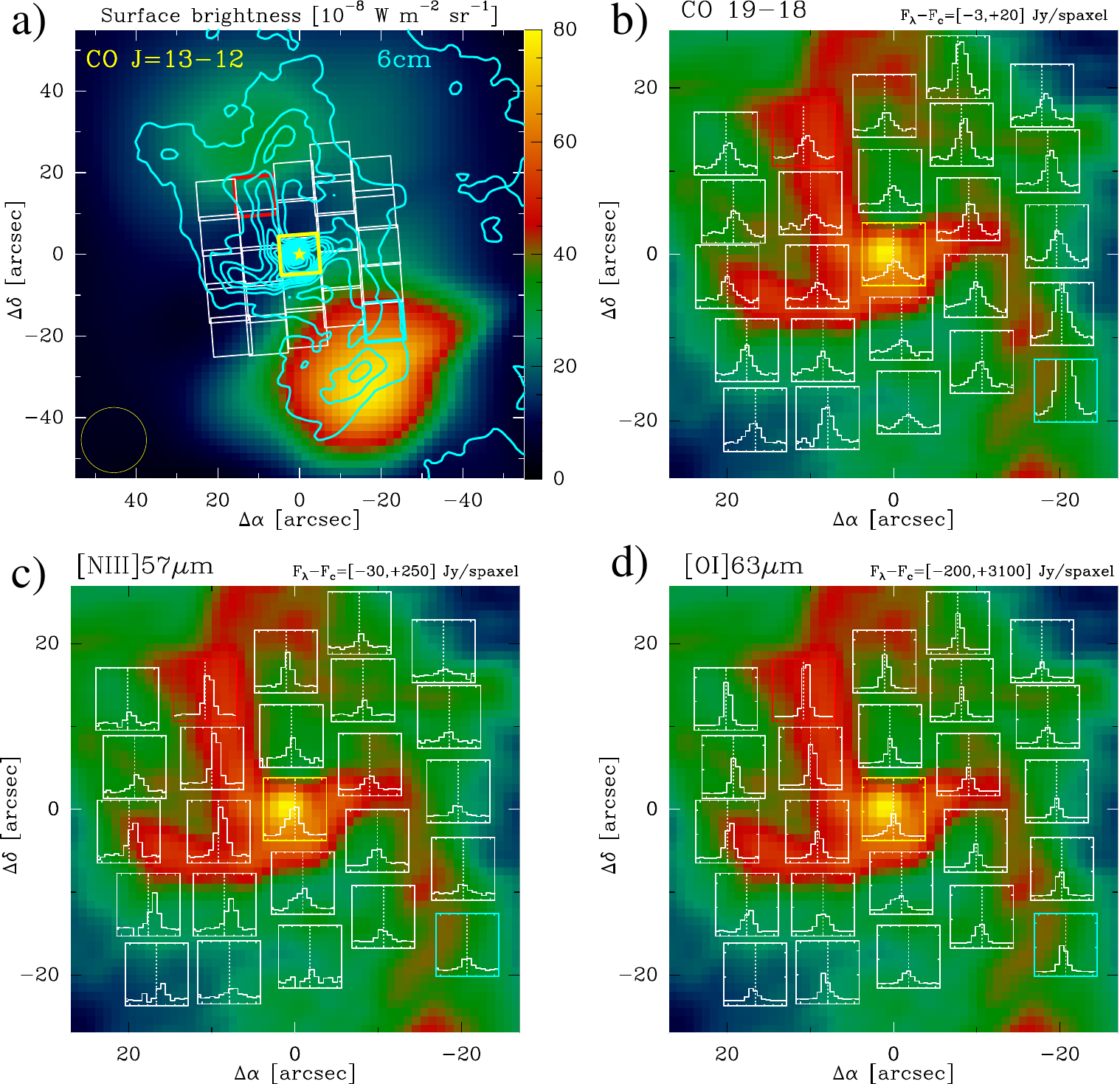} 
\caption{\textit{(a)} SPIRE-FTS $^{12}$CO $J$=13-12 sparse-sampling map of the GC and
VLA~6cm radio~continuum  showing ionized gas in the mini-spiral \citep[cyan contours,][]{Yus87}. 
Sgr\,A$^*$ is marked with a  star. The PACS footprint  is overplotted.
Panels \textit{(b), (c)} and \textit{(d)}: 
PACS~continuum-subtracted maps for the CO $J$=19-18 (137.196\,$\mu$m), [\NIII]\,57.317\,$\mu$m and [\OI]\,63.183\,$\mu$m lines.
The center of each spaxel corresponds to its offset position with respect to \Sgra.
The X-axis represents the $-$700 to $+$700 km\,s$^{-1}$ velocity scale. 
The line flux scale (Y-axis) is shown in each map.
The 6cm radio~continuum image  is shown in the background.}
\label{fig:show_area}
\end{figure*}

Figure~\ref{fig:show_area}$a$ shows the CO $J$=13-12 line intensity  map
of the central $\sim$4~pc of the Galaxy displaying 
extended excited~CO emission that peaks toward the northern and southern lobes of the CND
(hereafter $N$-CND and $S$-CND). 
Figures~\ref{fig:show_area}$b$, $c$ and $d$ show the CO $J$=19-18, [\NIII]\,57 and [\OI]\,63\,$\mu$m 
line spectral-maps respectively.

\begin{figure*}[ht]
\centering %
\includegraphics[width=1\textwidth,angle=0]{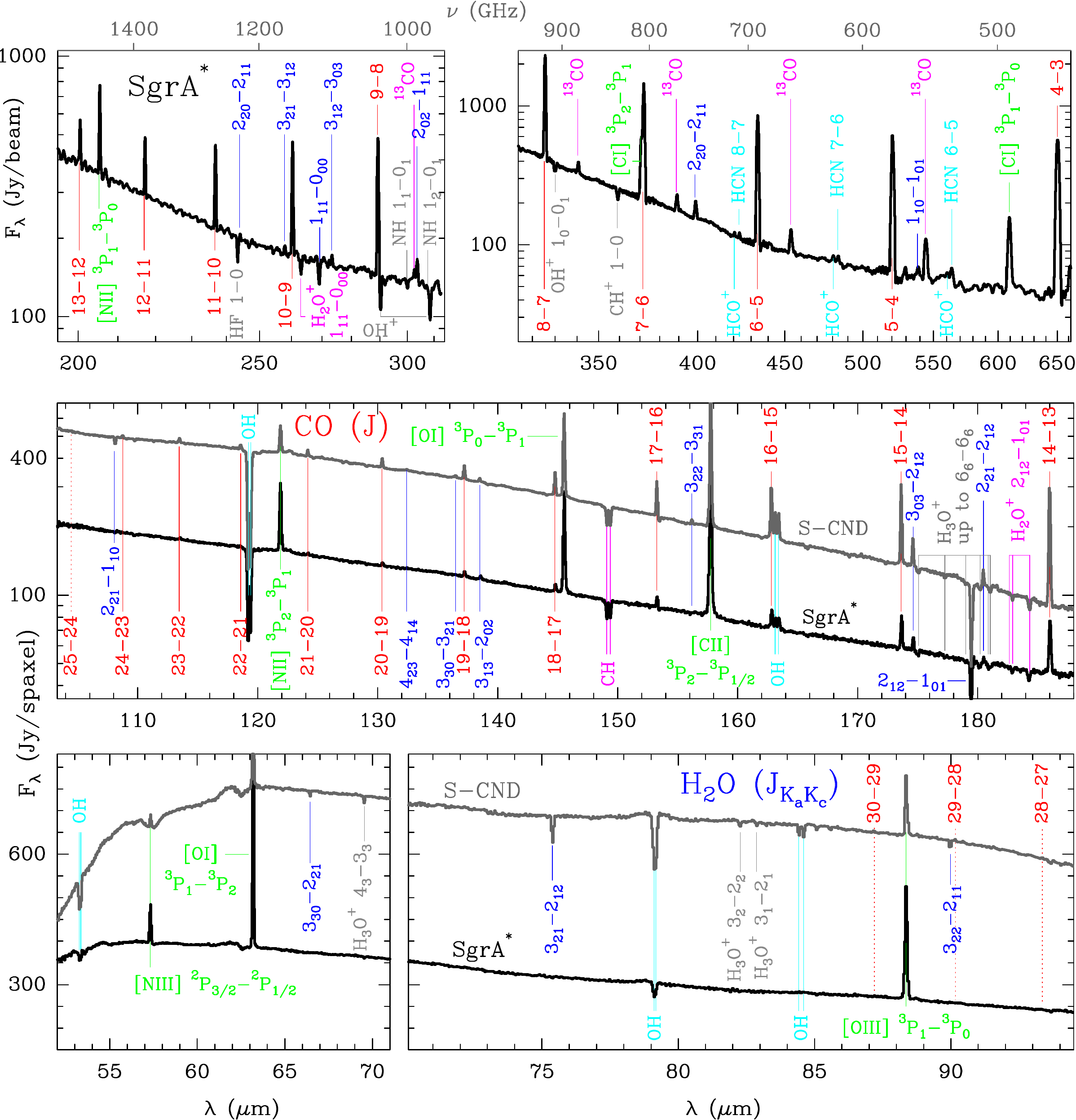} 
\caption{\textit{Top} panel: SPIRE-FTS spectrum toward Sgr\,A$^*$.
\textit{Middle} and \textit{bottom} panels: PACS spectra 
toward Sgr\,A$^*$ (black curves; yellow spaxel in Fig.~1$a$) and also
toward the $S$-CND (gray curves; blue spaxel in Fig.~1$a$).
Flux density units are  Jy\,spaxel$^{-1}$ for PACS and Jy\,beam$^{-1}$ for SPIRE.}
\label{fig:pacs_spire_spectra}
\end{figure*}

Figure \ref{fig:pacs_spire_spectra} shows the complete $\sim$52-671\,$\mu$m 
spectrum toward Sgr\,A$^*$ (black 
curves) and toward a bright  position at the inner edge of the  $S$-CND,  $\sim$1\,pc
 from \Sgra~(only for PACS; gray curves).  
The far-IR spectrum toward Sgr\,A$^*$ is dominated by  strong emission from 
atomic fine structure lines
([\OIII],  [\OI],   [\CII], [\NIII],  [\NII],  and [\CI]),
high-$J$ CO rotational lines (up to $J$=24-23 toward the central cavity and up to $J$=30-29 in the CND), mid-$J$ 
HCO$^+$ and HCN emission lines 
and  ground-state absorption lines from light hydrides (OH$^+$, H$_2$O$^+$, H$_3$O$^+$, CH$^+$, 
HF, CH,  NH, OH and  H$_2$O). Among the molecular ions, only H$_3$O$^+$ shows absorption lines
from excited levels (metastable levels up to $J_K$=$6_6$ toward the central cavity).
The detection of rotationally excited lines from H$_2$O, OH and H$_3$O$^+$ in absorption, 
suggests that a non negligible fraction of excited molecular gas resides at relatively 
low densities.

Owing to the $A_V$$\simeq$30 mag of extinction toward the GC \citep[][]{Gen10}, 
in the following
discussion we correct all line intensities ($I_{\rm 0}$=$C_{\lambda} I_{\rm obs}$) using an extrapolation of the  mid-IR extinction-law
derived by Lutz \cite{Lut99}  for \Sgra.
These corrections are $<$15\,\% in the far-IR (Table~1). 
The corrected line luminosities in the inner $\sim$30$''$$\times$30$''$ ($\sim$0.6\,pc in radius) 
are\footnote{We obtain $L_{\rm FIR}$(50-1000\,$\mu$m)$\simeq$10$^{5.1}$\,\Lsun~using
\textit{Herschel} photometric data from Etxaluze et al. (2011).}  $L_{\rm[OIII]}$$\simeq$885\,\Lsun, $L_{\rm[OI]}$$\simeq$855\,\Lsun, $L_{\rm[CII]}$$\simeq$230\,\Lsun,
$L_{\rm[NIII]}$$\simeq$130\,\Lsun, $L_{\rm[NII]}$$\simeq$120\,\Lsun, $L_{\rm ^{12}CO}$$\simeq$125\,\Lsun,
$L_{\rm ^{13}CO}$$\simeq$4\,\Lsun~and  $L_{\rm[CI]}$$\simeq$6\,\Lsun~(adopting $d$=8\,kpc).

\subsection{Ionized Gas and Neutral Atomic Gas}

The velocity resolution of PACS spectra at short wavelengths, $\lesssim$100\,km\,s$^{-1}$, allows one to resolve
shifts in the line profile peak velocity if they are large.  Figures~\ref{fig:show_area}$c$ and $d$ show the
[\OI]\,63 and [\NIII]\,57\,$\mu$m line maps.  
The Doppler shifts of both lines reveal motions of the neutral and of the ionized gas,
with blueshifted velocities observed toward the $S$-CND and redshifted toward the $N$-CND. 
The  [\NIII]\,57\,$\mu$m lines show a similar pattern to the  [\NeII]\,13\,$\mu$m  lines observed
at higher spatial and spectral resolution \citep{Iro12}. They are consistent with ionized gas streamers orbiting or falling
in a potential dominated by the central black hole \citep[][]{Ser85,Pau04}.
Note that both the  [\OI]\,63 and [\NIII]\,57\,$\mu$m lines show high-velocity wing emission 
($\gtrsim\pm$300\,km\,s$^{-1}$)
toward \Sgra~and narrower profiles at greater distances from the center. This high-velocity gas
is  likely associated with clouds/clumps moving fast and close to~\Sgra.

The spatial distribution of the [\NIII] lines follows
the dense filaments ($n_{\rm e}>10^{3.5}$\,cm$^{-3}$) of ionized gas orbiting 
\Sgra~\citep[the mini-spiral shown in Figure~1,][]{Yus87}.
This lower limit to the electron density ($n_{\rm e}$)  is estimated by comparing the 
[\OIII]\,88--to--[\OIII]\,52  line intensity ratio  of $0.21\pm0.12$ (hereafter [\OIII]\,88/52) 
with the semi-empirical 
prescription of Rubin et al.~\cite{Rub94}. 

The [\NIII]\,57 and [\NII]\,122\,$\mu$m  line intensities can also be used to estimate the
effective temperature of the ionizing radiation ($T_{\rm eff}$) for a given $n_{\rm e}$ (Rubin et al. 1994).
The [\NIII]\,57/[\NII]\,$122=1.31\pm0.51$ intensity ratio  toward 
\Sgra~results in $T_{\rm eff}$$\simeq$35,000$\pm$1000\,K \citep[see][for photoionization detailed models]{Shi94}. 

The strong UV radiation field from stars in the central parsec is thought to dominate the heating 
of the dust grains and  of the neutral atomic gas \citep{Gen85,Jac93}.
The low  ($L_{\rm[OI]}$+$L_{\rm[CII]}$)/$L_{\rm FIR}$$\simeq$8$\times$10$^{-3}$ luminosity ratio
toward the central cavity agrees with the low efficiency heating mechanisms expected in 
photodissociation regions (PDRs). 
In addition, 
[\OI]\,63/[\CII]\,$158=3.32\pm1.31$ and  [\OI]\,$145/63=0.12\pm0.05$ line intensity ratios are observed toward \Sgra.
These are   similar to the ratios observed in 
strongly irradiated PDRs like the Orion Bar \citep[][]{Ber12} but are significantly lower than the
expected ratios in shocked gas and in X-ray dominated regions   \citep[XDRs;][]{Mal96}. 
Therefore,
the neutral atomic gas ($n_{\rm H}$ $\sim$10$^{4-5}$\,cm$^{-3}$) is predominantly heated by UV photons
($G_0>10^{4}$ times the mean interstellar radiation field).
Atomic gas temperatures $\lesssim$500\,K are expected in this PDR range
\citep{Wol90}. Nevertheless, narrow (unresolved)  absorption components from foreground gas in the GC 
\citep[\textit{e.g.,}][]{Son13}  can affect the total [\OI]\,63 and [\CII]\,158\,$\mu$m  fluxes measured by PACS 
at medium spectral resolution (both are ground-state transitions).
Hence, the actual ratios may be different,
and our unresolved  [\OI] and  [\CII] intensities may have less diagnostic power.

\subsection{Hot Molecular Gas}

Figure~\ref{fig:show_co_diagrams}$a$ shows all detected $^{12}$CO    lines toward
the central cavity on a \textit{rotational diagram} that assumes extended emission.
The average $^{12}$CO/$^{13}$CO line intensity ratio is 22$\pm$9 (lines $J$=5-4 to 10-9),
thus consistent with the  $^{12}$C/$^{13}$C$\simeq$20-25 isotopic ratio inferred in the Sgr\,A complex  \citep{Pen80}.
Hence, the observed line intensity ratios are compatible with optically thin  $^{12}$CO line emission
(see the next section).

By fitting the  $^{12}$CO lines detected by SPIRE and PACS   independently,
we obtain $T_{\rm rot}$(SPIRE)$\simeq$87\,K  and $T_{\rm rot}$(PACS)$\simeq$232\,K respectively.
$T_{\rm rot}$ is a good measure of  the gas 
temperature only in the high density limit (close to \textit{local thermodynamic equilibrium}, LTE).
In this case, the two slopes could be associated with two different  temperature components,
the hotter one representing $\lesssim$4\,\% of the total  $^{12}$CO column density.
Alternatively, $T_{\rm rot}$  can reflect much higher gas temperatures
if the density is significantly lower than the critical density for collisional excitation
($n_{\rm cr}(\rm H_2)>10^7$\,cm$^{-3}$ for the  observed high-$J$ lines). 

In fact, a closer inspection of the rotational diagram shows that
$T_{\rm rot}$ increases with $J$, from $T_{\rm rot}$(14-18)\,$\simeq$160~K to $T_{\rm rot}$(19-24)\,$\simeq$250~K.
Therefore, the rotational diagram shows a moderate \textit{positive curvature} and thus even
a single temperature component in LTE cannot 
 explain the CO emission detected by PACS.
\cite{Neu12} pointed out that a moderate positive curvature can be explained either by (1) 
a \textit{subthermally excited} single temperature  component ($T_{\rm k} \gg T_{\rm rot}$) or (2) 
 multiple-temperature components (that could be in LTE). 

In order to constrain the range of physical conditions that  reproduce the observed CO intensities,  
we have run a grid of non-local, non-LTE isothermal models
\citep[][]{Goi06} using  $^{12}$CO-H$_2$ collisional rates from Yang et al. (2010).
We used a constant beam-averaged CO column density, $N$(CO), and a nonthermal velocity dispersion 
$\sigma=65$~km\,s$^{-1}$ (from turbulence and macroscopic gas motions),
implying  $\Delta v\simeq150$~km\,s$^{-1}$ line-widths. 
These broad widths are 
consistent with the mid-$J$ CO line-widths observed with \textit{Herschel}/HIFI (T.A.~Bell~2013, private communication).
We adopt $N$(CO)=$\chi$(CO)$\times$$N$(H$_2$)=10$^{18}$\,cm$^{-2}$ \citep{Gen85},
\textit{i.e.,}~we assume a CO abundance of $\simeq$10$^{-4}$ and take, from photometric measurements,
 $N$(H$_2$)$\simeq$10$^{22}$\,cm$^{-2}$ (A$_V$$\sim$10)
 in the central cavity \citep[][]{Etx11}.

Figure~\ref{fig:show_co_diagrams}$b$ shows the model results in the form of iso-$T_{\rm rot}$ contours.
To make this plot, we first created rotational diagrams from each model and determined $T_{\rm rot}$ by fitting
a straight line to the synthetic CO line intensities in the $J_{\rm up}$=16-24 range. 
Figure~\ref{fig:show_co_diagrams}$b$ shows that in terms of excitation alone,
the same $T_{\rm rot}$(PACS)$\approx$232\,K can be obtained for different combinations of density and temperature.
In a second step, we searched for the range of $n$(H$_2$) and $T_{\rm k}$ values that better reproduce the
observed $^{12}$CO lines by fitting their absolute intensities. 
For the adopted  $N$(CO), the best-fit parameters are obtained
around $T_{\rm k}$$\simeq$10$^{3.1}$\,K and $n$(H$_2$)$\simeq$10$^{3.7}$\,cm$^{-2}$. The resulting best-fit 
rotational diagram is shown as
a  green curve in Figure~\ref{fig:show_co_diagrams}$a$ (note that it is also 
consistent with the 3$\sigma$ upper limits for  higher-$J$  undetected  lines). 

In a third step, we compared the observations with the more extensive  model grid of Neufeld (2012) and studied the dependence of our results on
the assumed $N$(CO). For isothermal models, the gas temperature is well constrained and
$T_{\rm k}$$\simeq$10$^{3.1}$\,K reproduces the observed CO rotational ladder 
(also the lines detected by PACS alone)
independently of the assumed $N$(CO). 
Even lower  densities  (for less realistic higher $N$(CO) columns)  can also  reproduce the CO ladder.

In addition to the isothermal solution, multiple gas temperature components can also explain a CO rotational 
diagram with positive curvature.
\cite{Neu12} studied the case of a medium with a power-law distribution of temperatures, $dN$(CO)/$dT_{\rm k}=aT_{\rm k}^{-b}$.
Including all CO lines observed by \textit{Herschel} in the fit gives $b\simeq2.0-2.5$ and  $n$(H$_2$)$\simeq$10$^{4-5}$~cm$^{-3}$.
In other words, 
although a small fraction of the total $^{12}$CO column exists
at  $T_{\rm k}$$>$300\,K ($\lesssim$4$\%$), 
most  of the $N$(CO) column will be at lower temperatures and higher densities than those implied by the isothermal solution.
We therefore conclude that the observed CO lines are consistent with either a single,  \textit{hot} 
($T_{\rm k}$$\simeq$10$^{3.1}$\,K), low-density ($n$(H$_2$)$\lesssim$10$^4$\,cm$^{-3}$) 
component, or  with multiple, cooler components at a higher density. In the latter case, the required density will be above
the beam-averaged gas densities in the central cavity \citep[$\sim$10$^{3-4}$\,cm$^{-3}$;][]{Etx11}, implying  that
the \textit{hot molecular gas} in the vicinity of \Sgra~does not have a homogeneous distribution but 
fills a small fraction of the volume.
 
\section{Discussion}

\begin{figure*}[ht]
\centering %
\includegraphics[width=1\textwidth,angle=0]{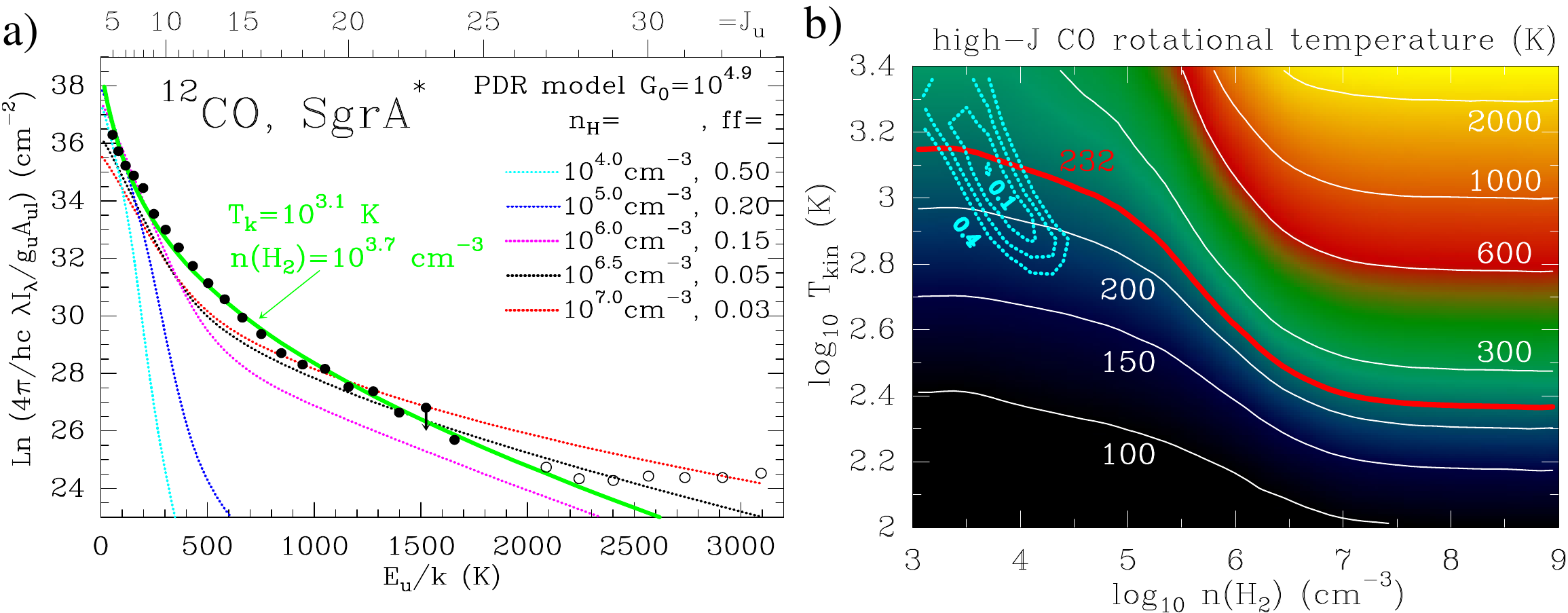} 
\caption{\textit{(a)} Observed $^{12}$CO~rotational~diagram for Sgr\,A$^*$ (filled circles).   
Open circles correspond to 3$\sigma$ upper limits for non-detections.
The green curve represents the best-fit isothermal model with $N$($^{12}$CO)=10$^{18}$~cm$^{-2}$.
Dashed curves show rotational diagrams from PDR models with different
 filling factors ($I_{\rm obs}$=ff$\cdot$$I_{\rm PDR}$). 
\textit{(b)} Synthetic CO rotational temperatures ($J_{\rm u}$=14-24 range) obtained  
from a grid of isothermal non-LTE models.
Cyan contours show the rms levels of  log$_{10}$($I_0$/$I_{\rm model}$) for fits to the absolute 
line intensities from  0.1 (best-fits with a rms error of $\sim$25\%) to 0.4 in rms error steps of 0.1.}
\label{fig:show_co_diagrams}
\end{figure*}

In this section we  discuss the possible heating mechanisms of the \textit{hot} molecular gas
toward the central parsec.
In order to evaluate the role of UV~radiation in the heating and  excitation of CO, 
we used an updated version of the \textit{Meudon} PDR code \citep{LeB12} to compute
synthetic CO rotational diagrams for the integrated CO emission from $A_{\rm V}$=0 to 10.
We adopted $G_0$=10$^{4.9}$ \citep{Wol90} and different gas densities.
Our photochemical model includes selective photodissociation of CO-isotopologues and $^{13}$C fractionation. 
For the considered range of densities, 
selective photodissociation slightly increases the $^{12}$CO/$^{13}$CO column density ratio   
over the $^{12}$C/$^{13}$C isotopic ratio (by $\lesssim$25\%)  at the $A_{\rm V}$$<$2 surface layers  
where CO columns are still low. 
Deeper inside, as the gas temperature decreases, $^{13}$C isotope exchange starts to be important
 and the  $^{12}$CO/$^{13}$CO column density 
ratio can  be lower than the  $^{12}$C/$^{13}$C  ratio. All in all, we conclude that the beam-averaged $N$($^{12}$CO) toward
\Sgra~cannot be much larger than the adopted $\sim$10$^{18}$\,cm$^{-2}$. 
For these columns and large velocity dispersions,
the observed  $^{12}$CO lines are optically thin.
Figure~\ref{fig:show_co_diagrams}$a$ shows the resulting CO diagrams for different PDR models and filling factors.
By comparing with observations, we see that low-density PDRs  ($n_{\rm H}$$\leq$10$^{5}$\,cm$^{-3}$)
are not able to reproduce the high-$J$ CO emission and an extra heating/excitation source is needed. 
In addition to photoelectric heating, denser PDRs ($n_{\rm H}$$\simeq$10$^{6-7}$\,cm$^{-3}$)  heat larger columns
 of molecular gas (to T$_k$$\sim$10$^3$\,K) by
vibrational heating from collisional deexcitation of
 UV-pumped H$_2$ molecules. Therefore, in addition to an extended low-density medium,
a small filling factor ensemble
 of irradiated dense  clumps/clouds could be responsible of the high-$J$ CO emission \citep{Bur90}.
Our best combined PDR models, however, do not provide an entirely  satisfactory fit of the CO~rotational ladder. 
This result is consistent with
the lack of good high-density fits to the high-$J$ CO lines (see previous the section)
and suggests that UV~radiation alone can not heat the 
\textit{hot} molecular gas.

Interestingly, the SPIRE-FTS spectrum toward \Sgra~resembles that of the M82 starburst galaxy \citep{Kam12}. However,
the \textit{hot} CO rotational temperatures inferred toward~\Sgra~are significantly higher than those seen in 
strongly irradiated PDRs like the Orion Bar, which shows a  rotational 
diagram (up to $J$=21-20) that can be fitted with a single $T_{\rm rot}$(CO)$\simeq$150\,K component
\citep[][C.~Joblin et al. in preparation]{Hab10}. 
In addition, even toward the strongly UV-irradiated central cavity,
the observed $L$(CO)/$L_{\rm FIR}$$\simeq$10$^{-3}$ luminosity ratio 
is higher than the expected ratio in PDRs and XDRs models \citep[][]{Mei13},
and it is indeed higher than the observed value in the Orion Bar  ($L$(CO)/$L_{\rm FIR}$$\simeq$3$\times$10$^{-4}$; C.~Joblin~2013, private communication).

The current X-ray luminosity near \Sgra~is  rather low 
\citep[$L_{\rm X}$(2-120\,keV)$< 10^{36}$~erg\,s$^{-1}$;][]{Bel06},
far lower than that expected from black hole accretion models.  Hence,  
any incident X-ray flux at a typical distance of $\sim$0.5\,pc from the source
  ($F_{\rm X}< 0.03$~erg\,cm$^{-2}$\,s$^{-1}$) would be too low to heat a significant fraction
of the molecular gas well above $T_{\rm k}$$\sim$100\,K
\citep[][]{Mal96}.

High cosmic-ray (CR) ionization rates ($\zeta_{\rm CR}\gtrsim10^{-15}$~~s$^{-1}$) have been inferred 
in the GC region from  H$_{3}^{+}$ observations, and even higher rates have been proposed for the vicinity of
\Sgra~\citep[$\sim$2$\times$10$^{-14}$~s$^{-1}$;][]{Got08}. At least qualitatively, our detection of H$_3$O$^+$ 
absorption lines from excited metastable levels indeed suggests that $\zeta_{\rm CR}$ may be high.
Nevertheless, the ionization fraction of the molecular gas in the central parsec seems lower
than that in much more  extreme X-ray dominated AGNs like Mrk~231 
($L_{\rm X}$(2-10\,keV)$\simeq 6\times10^{43}$~erg\,s$^{-1}$), where strong CH$^+$, OH$^+$ and H$_2$O$^+$ emission lines
have been detected \citep{Wer10}. It also has to be  lower than in the ULIRG galaxy Arp~220, 
where the detection of 
many excited OH$^+$ and H$_2$O$^+$ absorption lines has been associated with very enhanced X-ray/CR 
ionization rates \citep[$\zeta_{\rm X,CR} > 10^{-13}$~s$^{-1}$;][]{Gon13}.
Our observations toward \Sgra~show that  OH$^+$ and H$_2$O$^+$  only produce appreciable ground-state absorption lines,
and they are known to arise from semi-atomic diffuse clouds, where their
columns  are proportional to $\zeta_{\rm CR}$ \citep{Ger10,Neu10,Hol12}.
Hence, $\zeta_{\rm CR}$ toward \Sgra~is very  likely higher than in Galactic disk clouds, but lower than $\zeta_{\rm X,CR}$ in
Mrk~231 or Arp~220. Simple thermodynamic considerations show that even $\zeta_{\rm CR}$=2$\times$10$^{-14}$~s$^{-1}$
would only heat the gas to a few tens of K \citep[see also][]{Glas12}. 
Therefore,  neither X-rays nor CRs presently  dominate the heating of the \textit{hot} molecular gas
near \Sgra.

Low-density shocks (and related supersonic turbulence dissipation and magnetic viscous heating) are promising candidates
in the highly magnetized GC environment \citep{Mor96}.
In particular, non-dissociative, magnetohydrodynamic shocks  tend to produce regions that are roughly isothermal, 
reaching very high temperatures without destroying molecules \citep[$T_{\rm k}$$>$1000~K for shock velocities $v_{\rm s}$$>$20~km\,s$^{-1}$ 
in C-type shock models by][]{Kau96}. 
In fact, shocks with a variety of densities, $v_{\rm s}$ and magnetic field strengths dominate the heating of the  hot molecular gas seen in
protostellar outflows \citep[with $L$(CO)/$L_{\rm FIR}$$\simeq$2$\times$10$^{-3}$ observed in Serpens~SMM1;][]{Goi12}.
Hence, the \textit{hot} CO gas inferred toward \Sgra, the high $L$(CO)/$L_{\rm FIR}$ ratio and
 the almost thermal H$_2$ rovibrational spectrum \citep[][]{Tan89} suggest that, 
in addition to UV-driven excitation, shocks  \textit{contribute} to
the heating of the \textit{hot} molecular gas in the nucleus of the Galaxy.
Indeed, if a small filling factor ensemble of dense clumps/clouds does not exist, 
shocks likely \textit{dominate}.

Whether the required  shocks are produced within high-velocity molecular gas falling toward
the very center region  \citep[][]{Gil12}, arise in clump-clump collisions \citep[][]{Mar97} or 
in outflows driven by high-velocity stellar winds \citep{Naj97} or
by protostars in the central~parsec \citep{Nis13} is still uncertain.

\acknowledgments

We thank C.~Lang and M.A.~Requena-Torres for providing us with the VLA 6\,cm radiocontinuum image
in CLASS format, and C.~Joblin and O.~Bern\'e for useful discussions on  the Orion Bar.
We thank the Spanish MINECO for funding support from grants AYA2009-07304, CSD2009-00038
and S2009ESP-1496, and NASA through an award issued by JPL/Caltech.
J.R.G. is supported by a \textit{Ram\'on y Cajal}  research contract.




\clearpage

\begin{deluxetable}{cccccc}
\tabletypesize{\scriptsize} 
\tabletypesize{\tiny} 
\tablecaption{Atomic and CO  Line Intensities Toward \Sgra\label{tbl-fluxes}}
\tablewidth{0pt}
\tablehead{
\colhead{Species} & \colhead{Transition} & \colhead{$\lambda$($\mu$m)}  & \colhead{$E_{\rm u}$/$k$\,(K)} &
 \colhead{$I_{\rm obs}$$^a$} & $C_{\lambda}$$^b$}
\startdata
$[$\OIII$]$ &  $^3P_2-{^3P}_1$        &  51.815     & 441     & 1.52E-05$^c$ & 1.152\\
$[$\NIII$]$ &  $^2P_{3/2}-{^2P}_{1/2}$ &  57.317     & 251     & 2.69E-06     & 1.132\\
$[$\OI$]$   &  $^3P_1-{^3P}_2$        &  63.184     & 228     & 1.62E-05     & 1.116\\
$[$\OIII$]$ &  $^3P_1-{^3P}_0$        &  88.356     & 163     & 3.30E-06     & 1.076\\
$[$\NII$]$  &  $^3P_2-{^3P}_1$         & 121.898    & 188     & 2.21E-06     & 1.051\\
$[$\OI$]$   &  $^3P_0-{^3P}_1$        & 145.525     & 327     & 2.11E-06     & 1.041\\
$[$\CII$]$  &  $^2P_{3/2}-{^2P}_{1/2}$ & 157.741     & 91      & 5.26E-06     & 1.037 \\
$[$\NII$]$  &  $^3P_1-{^3P}_0$        & 205.178     & 70      & 4.51E-07     & 1.026\\
$[$\CI$]$   &  $^3P_2-{^3P}_1$        & 370.414     & 63      & 1.22E-07     & 1.013\\
$[$\CI$]$   &  $^3P_1-{^3P}_0$        & 609.133     & 24      & 1.95E-08     & 1.007\\\hline
$^{12}$CO    &  $J$=24-23             & 108.763     & 1656.6  & 1.40E-08     & 1.058\\
$^{12}$CO    &  $J$=23-22             & 113.458     & 1524.3  & 3.49E-08$^d$ & 1.055\\
$^{12}$CO    &  $J$=22-21             & 118.581     & 1397.4  & 2.37E-08     & 1.052\\
$^{12}$CO    &  $J$=21-20             & 124.193     & 1276.1  & 3.95E-08     & 1.049\\
$^{12}$CO    &  $J$=20-19             & 130.369     & 1160.3  & 3.58E-08     & 1.047\\
$^{12}$CO    &  $J$=19-18             & 137.196     & 1049.9  & 5.26E-08     & 1.044\\
$^{12}$CO    &  $J$=18-17             & 144.784     &  945.0  & 4.67E-08     & 1.041\\
$^{12}$CO    &  $J$=17-16             & 153.267     &  845.6  & 5.23E-08     & 1.038\\
$^{12}$CO    &  $J$=16-15             & 162.812     &  751.8  & 7.54E-08     & 1.035\\
$^{12}$CO    &  $J$=15-14             & 173.631     &  663.4  & 9.66E-08     & 1.032\\
$^{12}$CO    &  $J$=14-13             & 185.999     &  580.5  & 1.30E-07     & 1.030\\
$^{12}$CO    &  $J$=13-12             & 200.272     & 503.2   & 1.59E-07     & 1.027\\
$^{12}$CO    &  $J$=12-11             & 216.927     & 431.3   & 1.94E-07     & 1.025\\
$^{12}$CO    &  $J$=11-10             & 236.613     & 365.0   & 2.38E-07     & 1.022\\
$^{12}$CO    &  $J$=10-9              & 260.240     & 304.2   & 2.75E-07     & 1.020\\
$^{12}$CO    &  $J$=9-8               & 289.120     & 248.9   & 2.84E-07     & 1.017\\
$^{12}$CO    &  $J$=8-7               & 325.225     & 199.1   & 3.89E-07     & 1.015\\
$^{12}$CO    &  $J$=7-6               & 371.650     & 154.9   & 3.08E-07     & 1.013\\
$^{12}$CO    &  $J$=6-5               & 433.556     & 116.2   & 2.02E-07     & 1.011\\
$^{12}$CO    &  $J$=5-4               & 520.231     & 83.0    & 1.34E-07     & 1.009\\
$^{12}$CO    &  $J$=4-3               & 650.252     & 55.3    & 7.81E-08     & 1.007\\\hline
$^{13}$CO    &  $J$=12-11             & 226.898     & 412.4   & 6.53E-09     & 1.023\\
$^{13}$CO    &  $J$=11-10             & 247.490     & 348.9   & 6.68E-09     & 1.021\\
$^{13}$CO    &  $J$=10-9              & 272.205     & 290.8   & 8.52E-09     & 1.019\\
$^{13}$CO    &  $J$=9-8               & 302.415     & 237.9   & 1.94E-08     & 1.016\\
$^{13}$CO    &  $J$=8-7               & 340.181     & 190.4   & 2.00E-08     & 1.014\\
$^{13}$CO    &  $J$=7-6               & 388.743     & 148.1   & 1.63E-08     & 1.012\\
$^{13}$CO    &  $J$=6-5               & 453.498     & 111.1   & 6.25E-09     & 1.010\\
$^{13}$CO    &  $J$=5-4               & 544.161     & 79.3    & 1.08E-08     & 1.008\\\hline
\enddata
\tablenotetext{a}{Observed lines intensities above~3$\sigma$ in W\,m$^{-2}$\,sr$^{-1}$.
Absolute calibration accuracy up to $\sim$30$\%$.}
\tablenotetext{b}{Extinction correction factors. $^c$From Shields \& Ferland (1994).}
\tablenotetext{d}{Blended with the  $o$-H$_2$O 4$_{14}$-3$_{03}$  line.}
\end{deluxetable}

\clearpage

\end{document}